# Facile hBN–hBN Interfacial Overlap Engineering for Enhanced Quantum Emitter Formation

Nhat Minh Nguyen [1], Trung Vuong Doan [1], Md Shakhawath Hossain [1], Akila Elangasinghe [2], Duc Anh Ngo [1], Ha Ngoc Duy Huynh [1], Thi Ngoc Anh Mai [1], Yongliang Chen [3], Kenji Watanabe [4], Takashi Taniguchi [5], Michael G. Ruppert [2], Chaohao Chen [1], Xiaoxue Xu [1], Toan Dinh [6,7], and Toan Trong Tran [1, †]

[1] School of Electrical, Mechanical and Biomedical Engineering, University of Technology Sydney, Ultimo, NSW, 2007, Australia

[2] Centre for Audio, Acoustics and Vibration, University of Technology Sydney, Ultimo, NSW 2007, Australia

[3] Department of Physics, The University of Hong Kong, Pokfulam, Hong Kong, China

[4] Research Center for Electronic and Optical Materials, National Institute for Materials Science, 1-1 Namiki, Tsukuba 305-0044, Japan

[5] Research Center for Materials Nanoarchitectonics, National Institute for Materials Science, 1-1 Namiki, Tsukuba 305-0044, Japan

[6] School of Engineering, University of Southern Queensland, Toowoomba, Queensland 4350, Australia

[7] Center for Future Materials, University of Southern Queensland, Toowoomba, Queensland 4350, Australia

[†] Corresponding author: Toan Trong Tran | trongtoan.tran@uts.edu.au

## Abstract

Quantum emitters in two-dimensional materials, particularly hBN, are promising platforms for quantum technologies. However, achieving high-density emitters at predetermined locations while preserving optical quality remains challenging. Here, we introduce a facile, cost-effective double-layer all-dry transfer approach to deterministically create overlap regions between hBN flakes. These pre-defined capped regions exhibit a significantly enhanced emitter density, with up to a 15-fold increase compared to uncapped areas. Importantly, this method does not compromise emitter quality: emitters within overlap regions demonstrate excellent optical performance, including high signal-to-background and signal-to-noise ratios, large Debye–Waller factors, high brightness, and strong spectral stability. Possible defect configurations are also discussed to contextualize the observed emission characteristics. This scalable strategy enables preferential formation of quantum emitters in targeted regions, achieving higher densities than simple treatments such as plasma irradiation while avoiding the complexity of advanced fabrication techniques. The approach provides a practical pathway for integrating high-quality quantum emitters into scalable quantum photonic platforms.

## 1. Introduction

Solid-state quantum emitters (QEs) have become building blocks in quantum technology with promising applications in communication, computing, sensing, and information processing.[1] Among typical solid-state platforms such as quantum dots (InP/ZnSe/ZnS, $CsPbBr_3$),[2] upconversion nanoparticles ($Er^{3+}$ ion-doped $NaYF_4$),[3] and micro- or nanodiamonds,[4] hexagonal boron nitride (hBN) has gained significant attention due to its unique features.[5] Owing to its wide bandgap of ~6 eV, hBN hosts various optically active centers, e.g., vacancies, antisites, impurities, or complexes, which exhibit bright emission and high chemical and photostability.[6] Additionally, recent investigations indicate that some color centers in hBN are spin-addressable, demonstrating the immense potential to expand the range of quantum applications.[7, 8]

Since the first discovery in 2016,[9] substantial studies have focused on the fabrication and engineering of hBN QEs. In this context, an ideal emitter engineering process should deliver (i) high-density or pre-defined/localized emitters[10, 11] and (ii) emitters with high optical quality, characterized by low fluorescence background, reliable emission signal, and good stability.[12-14] Although various approaches have been widely studied, they still face persistent challenges. For instance, heavy-ion irradiation,[15] focused ion or electron beam,[16, 17] mechanical indentation,[18, 19] being invasive in nature, often require demanding experimental conditions (e.g., high vacuum) and sophisticated instrumentation (e.g. focus ion beam system). The resulting emitters, however, do not consistently exhibit the desired optical properties. For example, emitters produced by heavy ions (Ta, Ge) or by local stress exhibit substantial fluorescence background, broad linewidth (~40 nm), and blinking.[15, 18] On the contrary, global approaches such as annealing, with or without plasma pretreatment, generally yield limited emitter densities. Emitter density achieved by plasma-assisted and post-annealing processes, for instance, is still relatively low in recent studies.[20, 21] Furthermore, important characteristics of emitters such as photostability, signal-to-background (SBR), and signal-to-noise (SNR) ratio are still insufficiently reported, making it difficult to assess their suitability for applications that demand reliable optical performance (e.g., sensing and photonics). In addition, the microscopic origin of many hBN emitters is rarely discussed in studies focusing on emitter engineering. For these reasons, the engineering of emitters via a simple, low-cost, spatially defined strategy that maintains high spectral quality and stability, together with a more thorough discussion of their possible origins, remains of great importance.

In the present study, we demonstrate a straightforward, region-defined, and reproducible approach for engineering QEs using deterministic transfer of two-dimensional hBN flakes. Specifically, an hBN flake was transferred onto a pre-exfoliated hBN flake to form a localized overlap region, which was then oxygen-annealed. Our results indicate that the emitter density in the pre-defined overlap geometry is significantly higher than that of the uncapped (control) regions of the same flake. An areal density enhancement of up to 15.8× was observed, and the overlap regions still exhibited higher densities even after normalization for thickness, suggesting that, beyond the simple increase in the investigated area and thickness, there may be an additional contribution associated with our capping geometry. Furthermore, by providing a statistical analysis of spectral quality metrics, we demonstrate that our approach does not introduce any trade-off in the emitter quality, as the SBR, SNR, and Debye-Waller (DW) factor of emitters in overlap regions remain competitive to those in the control regions. Our approach offers a practical route to increase QE density in predefined regions without relying on complex nanofabrication or compromising emitter quality.

## 2. Results and discussion

The facile, pre-defined transfer of hBN flakes was performed using the lab-built setup shown in **Figure S1a (Supporting Information 1)**. Briefly, the substrate containing the pre-selected hBN flake was placed on stage A, while the glass carrying the PDMS stamp with another hBN flake was mounted on stage B. The transfer process was conducted by gently adjusting the x-, y-, and z- knobs of both stages with the aid of an optical microscope, without applying any external heating.

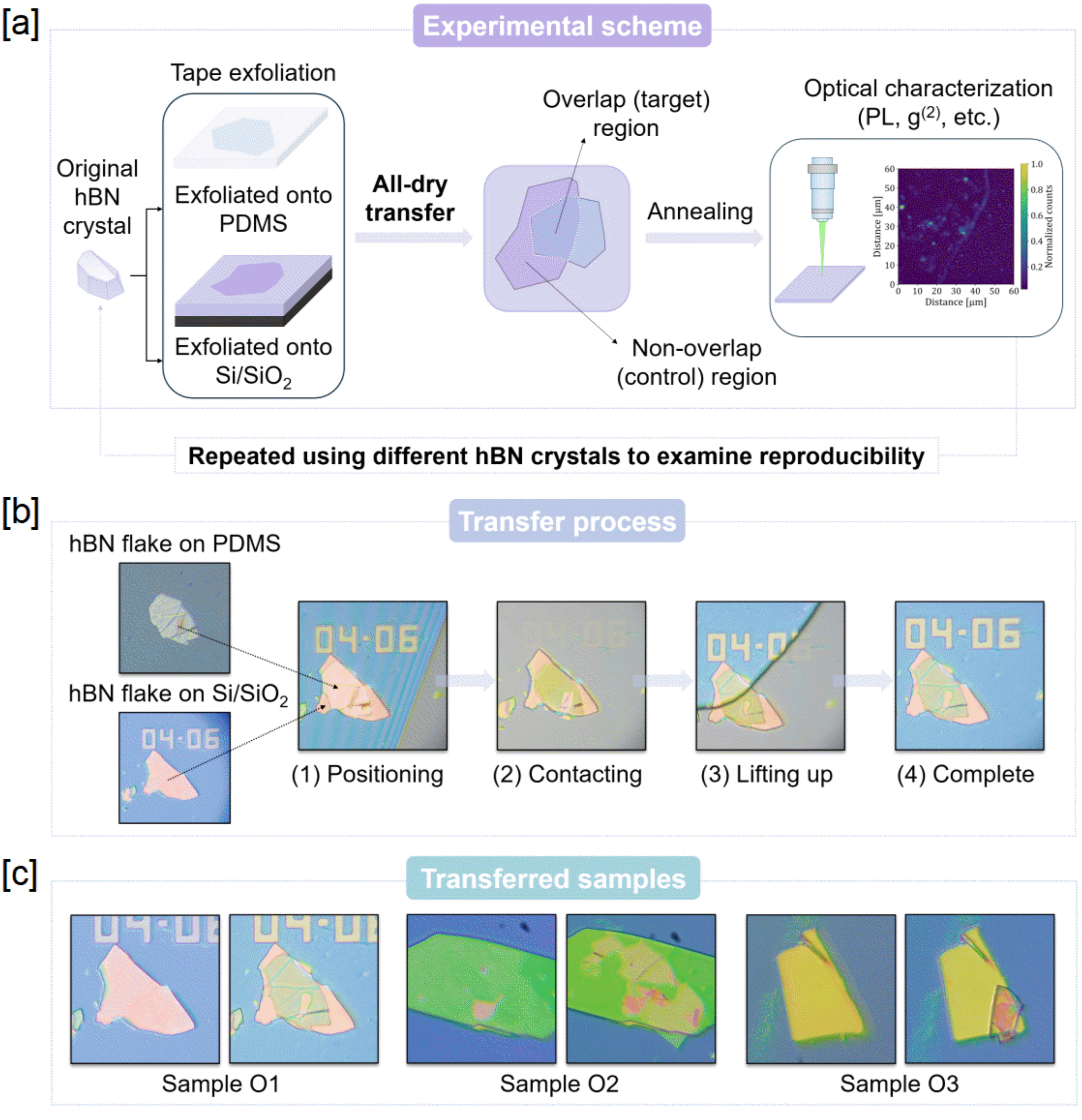

**Figure 1** | Experimental process. (a) Schematic illustration of the experiment conducted in this study (the overlap region is defined as the target region, while the non-overlap one serves as the control), (b) Optical images of a representative step-by-step all-dry transfer process, and (c) Optical images of the investigated samples before and after transfer. The full-scale images of these flakes with a scale bar are presented in the **Figure 2a** and **Figure S2a-b** – **Supporting Information 2**.

**Figure 1a** depicts the experimental process applied in this study. To assess the reproducibility of our method, the exfoliation, transfer, and annealing processes were independently repeated using different original hBN crystals. A representative step-by-step example of the transfer process is shown in **Figure 1b**. Initially, two targeted flakes on the substrate and the PDMS stamp were pre-selected and aligned relative to each other (step 1). The glass slide was then lowered gradually until the two flakes came into contact (step 2). After 5 minutes of contact, the glass slide was very gently lifted to complete the transfer, resulting in the overlapped region (steps 3 and 4). Optical images of samples O1, O2, and O3 before and after transfer are presented in Figure 1c, showing the successful, facile all-dry transfer processes. Noted that these optical images were taken immediately after transfer, prior to the annealing procedure. In this study, high-purity hBN synthesized via high-pressure, high-temperature (HPHT) technique was intentionally used to reduce the influence of uncontrolled impurities, which are often more prominent in commercial hBN sources, thereby allowing the role of the hBN–hBN overlap/capping geometry to be more clearly assessed.

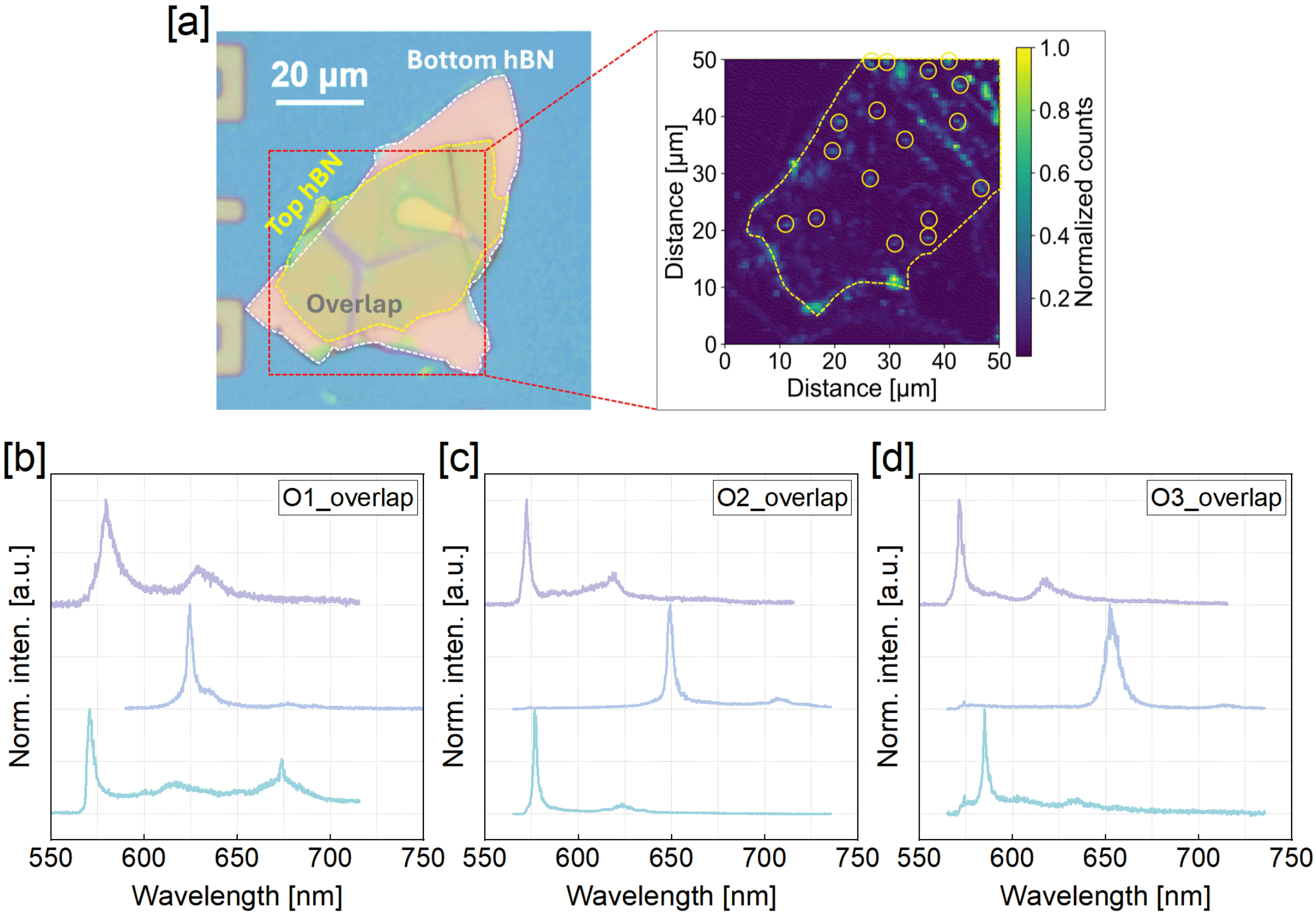

**Figure 2** | Optical measurements. (a) Optical image after annealing and corresponding confocal map of the overlap region in sample O1. (b), (c), and (d) Photoluminescence (PL) spectra of representative emitters in overlap regions of samples O1, O2, and O3, respectively. Optical images and confocal PL maps of sample O2, O3 are shown in Figure S2a-b, and PL spectra of representative emitters in control areas are presented in **Figure S3a-c**. In the confocal maps, the yellow circles indicate the position where PL spectra were collected. All the emitters found outside these capped areas are classified as control-region ones

After being annealed at 1000 °C in 200 sccm $O_2$, confocal scanning on both overlap and control regions of the resulting samples (O1, O2, and O3, corresponding to optical images shown in **Figure 1c**) was performed using the lab-built optical setup (refer to **Figure S1b – Supporting Information 1** for a schematic illustration). **Figure 2a** exhibits the optical images after annealing and the corresponding confocal PL map of the overlap region in sample O1, while those images and maps for samples O2 and O3 can be viewed in **Figure S2a-b**. Confocal maps reveal a relatively dense distribution of emitters within the overlap regions, whereas emitters were less frequently observed in the remaining uncapped (control) regions of the same flakes. This is reflected by the fraction analysis (**Figure S3**), where a higher fraction of emitters was observed in the overlap regions for all samples.

PL spectra of representative emitters from overlap regions of samples O1-O3 are displayed in **Figure 2b-c**, and those of emitters in control areas are presented in **Figure S4a-c**. In general, emitters were observed across all samples, regardless of region, and most exhibited negligible fluorescence background. This indicates that high-temperature annealing in oxygen can effectively produce optically active centers while simultaneously suppressing the fluorescence contribution from organic residues remaining after exfoliation and the subsequent transfer process.[11, 12] It is noteworthy that although most emitters displayed spectra with clear ZPLs accompanied by phonon sidebands (PSBs), some in control areas appeared with multi-peak profiles and relatively noisy signals. This suggests that certain color centers may be more sensitive to annealing, since 1000 °C in oxygen is highly oxidative. Similar spectra are also observed in recent studies that use oxygen annealing as an emitter-generation approach.[7, 22] However, this phenomenon was more frequently observed in control areas, whereas the emitters in overlap generally exhibited a clearer emission profile. Our observation suggests that the overlap geometry may help preserve some color centers during such an annealing process, as discussed later.

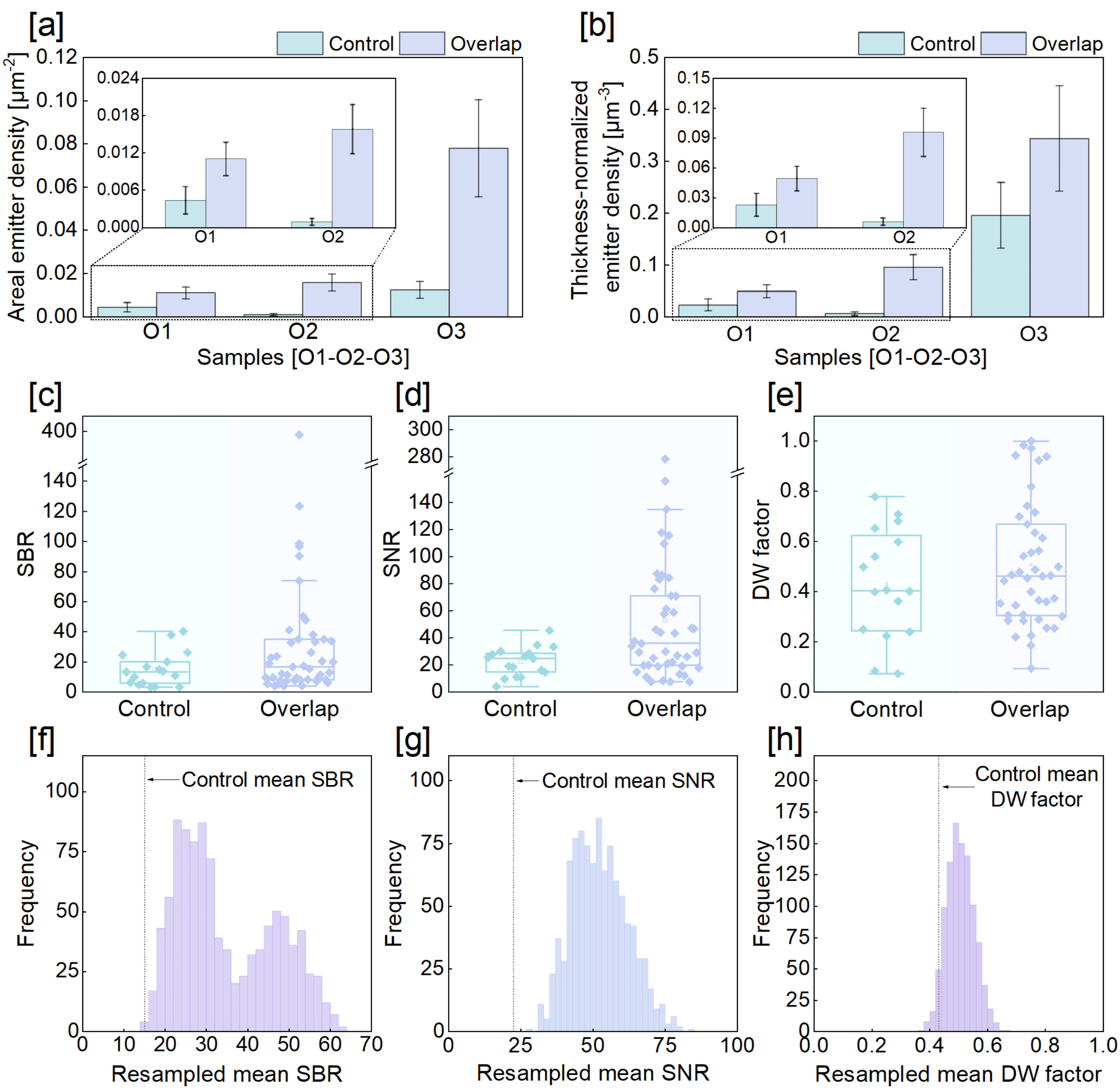

**Figure 3** | Statistics of emitter density and various quality metrics for overlap and control regions. (a) Areal and (b) Thickness-normalized emitter density for overlap and control regions. Statistical distribution of (c) SBR, (d) SNR, and (e) Debye-Waller (DW) factor for emitters in both regions across all samples. Monte Carlo random subsampling analysis comparing the resampled mean values of (f) SBR, (g) SNR, and (h) DW factor.

To evaluate the effectiveness of our approach, we conducted statistical analyses of density and quality for all emitters in both regions, as shown in **Figure 3**. Firstly, the areal emitter density—defined as the number of observed emitters divided by the total scanned area—for both regions in three samples was calculated (cf. **Figure S5a-c** and **Table S1 – Supporting Information 3** for details of the calculation).[11] As seen in **Figure 3a**, the areal emitter density within overlap is higher than that of control areas, being approximately 2.52, 15.8, and 6.29 times greater for samples O1, O2, and O3, respectively. The sample-to-sample variation may arise from differences in crystallinity and purity, which are typical between crystals during the growth process, since three samples were fabricated from different hBN bulk crystals.[23] Our approach enables the consistent identification of emitters in every overlap area, whereas some flakes treated with plasma and post annealing, as previously reported, do not host any centers.[21] These results suggest that the enhanced density cannot be solely explained by a simple addition of a hBN flake in the overlap region. In fact, if the increase originated only from the presence of an additional flake, the areal density would be expected to improve by approximately two-fold, considering that the top and bottom flakes were exfoliated from the same original crystal. However, the experimentally observed enhancement exceeds this simple expectation, from 2.52 to 15.8. We further investigate the effect of increased thickness in overlap regions on the enhanced emitter density. From the average flake thicknesses extracted from the AFM images, thickness-normalized emitter density was calculated (details can be viewed in **Figure S5a-c** and **Table S2 – Supporting Information 3**). If the improved emitter density was thickness-driven, the thickness-normalized density would be expected to be comparable between the overlap and the control regions. However, as shown in **Figure 3b**, the thickness-normalized emitter density of overlap regions remains higher than that of the control ones across all samples. Together with the areal emitter density analysis, these results suggest that the observed enhancement cannot be explained solely by either additional flake stacking or increased thickness. Therefore, an additional factor associated with the overlap geometry may contribute to the improved emitter density. Given the use of high-purity HPHT hBN with reduced and better-controlled impurity concentrations, which helps minimize the contribution of uncontrolled impurity-rich regions that can be more prominent in commercial hBN sources, the consistently higher areal and thickness-normalized emitter densities observed in the overlap regions support an important role of the hBN–hBN overlap/capping geometry in promoting or preserving optically active defects. Therefore, we tentatively attribute this to the preservation and activation of certain defects in the overlap region, which may either pre-exist in the hBN flakes or be transiently generated during high-temperature oxygen annealing. Although such defects may distribute across the entire flakes, those located in the overlap region may be better preserved during prolonged annealing, whereas their uncapped counterparts are likely to be less stable. Such preservation may therefore introduce an additional factor associated with the observed density enhancement.

To demonstrate that our overlap approach does not introduce a significant trade-off in emitter quality, we calculated several spectral quality metrics, including SBR, SNR, and DW factor, for all emitters in both the overlap and control regions (details of the calculations are presented in **Supporting Information S4)**. In this study, we considered a spot to be an emitter if it exhibited an SNR value $\geq 3$, the minimum threshold for reliable signal detection for further consideration,[24] and across three samples, we found a total number of 45 emitters in the overlap areas and 17 emitters across the control regions. From **Figure 3c-d**, the average values (Mean ± SD) of SBR and SNR are $15.10 \pm 11.54$ and $22.49 \pm 10.76$ for emitters in non-overlap regions, respectively, versus $35.12 \pm 62.06$ and $52.72 \pm 50.46$ for those in overlap regions, respectively. In general, emitters in overlap areas tend to exhibit higher brightness (as indicated by SBR) and more reliable signals (as indicated by SNR) compared to those in control areas. For DW factor, emitters in both regions exhibit an average value above 0.4, which is relatively reasonable at room temperature. The average value of those in overlap regions also appears to be higher ($0.51 \pm 0.25$ versus $0.43 \pm 0.22$). In all cases, the large deviation reflects the fact that several emitters possess exceptionally high spectral quality (i.e., SBR~400, SNR~280, and DW~1).

Considering the difference in the total number of emitters observed in the two regions, which may introduce uncertainty when comparing the spectral quality metrics, we performed a Monte Carlo subsampling analysis without replacement. Specifically, for each metric, 17 emitters in the overlap region were randomly selected to match the number of control-region emitters, and the mean values of their SBR, SNR, and DW factor were calculated. The calculation was repeated 1000 times, and the resulting distribution of the resampled mean values is presented as histograms shown in **Figure 3f-h**. Apparently, the resampled mean values of all quality metrics are predominantly shifted toward higher values than those of control-region emitters, suggesting that the overlap-region emitters exhibit comparable or even higher spectral quality than their counterparts in the control regions.

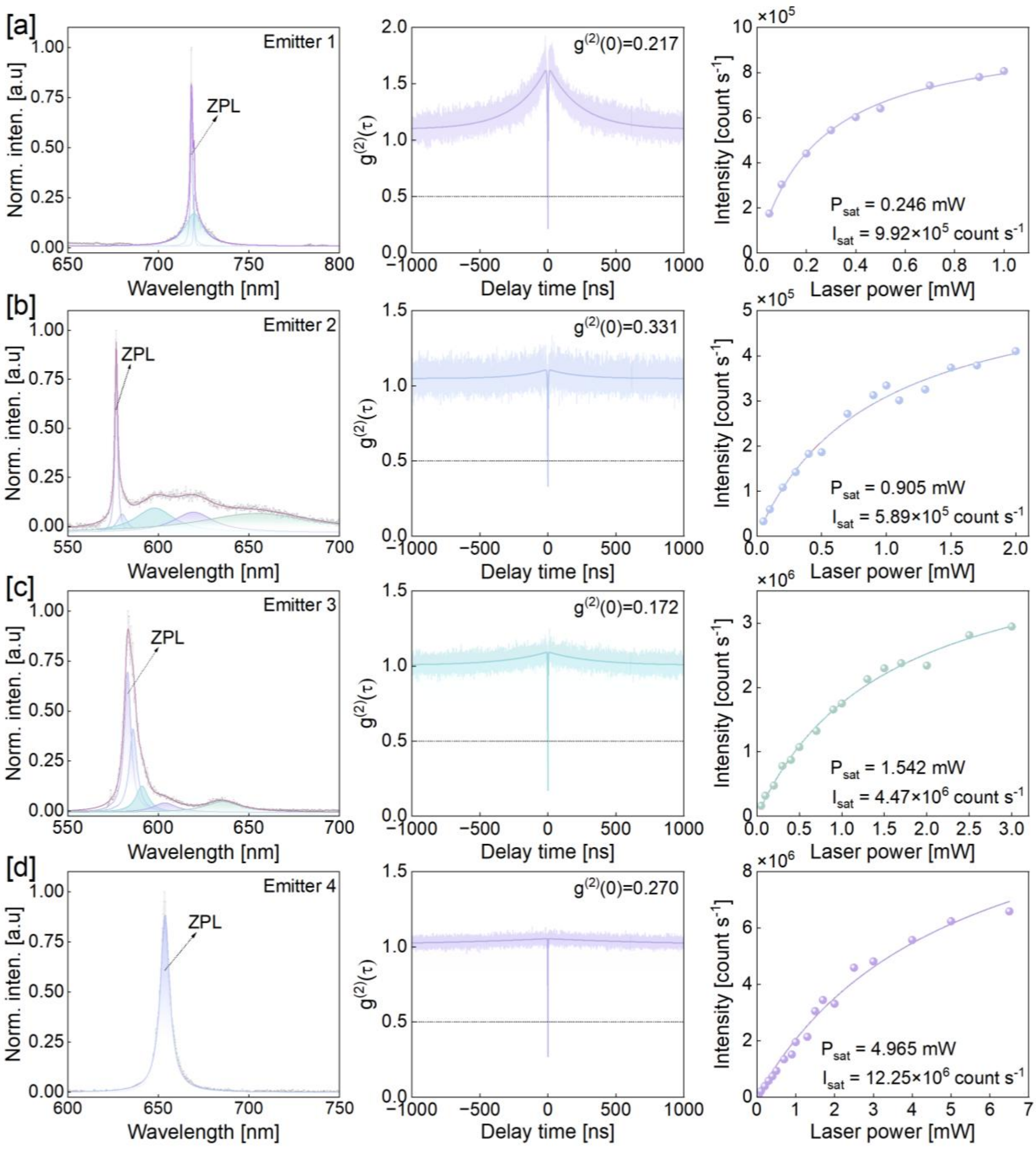

**Figure 4** | Characteristics of typical single quantum emitters in the overlap areas. Photoluminescence spectra with Lorentzian fitting (left panels), second-order autocorrelation measurement with fitting curves (middle panels), and fluorescence intensity versus laser power with fitting function (right panels) of (a) Emitter 1, (b) Emitter 2, (c) Emitter 3, and (d) Emitter 4

To further verify that the overlap approach can effectively enhance emitter density without compromising their optical performance, we characterized representative emitters from the overlap regions of samples O1-O3. The left panels of Figure 4a-d show the PL spectra of the four emitters with Lorentzian fits. These emitters exhibit clear, sharp ZPLs, accompanied by several PSBs, which are widely attributed to acoustic and optical phonon modes.[23, 25] The presence of fluorescence background or noise in these spectra is negligible, consistent with the effect of oxygen annealing, as discussed above. Notably, the full width at half maximum (FWHM) values of Emitter 1 and Emitter 2 were estimated at 1.24 nm and 1.78 nm, respectively, generally indicating reduced spectral broadening and better optical quality. Additionally, we noted that for Emitter 4, the Debye-Waller (DW) factor—representing the contribution of ZPL to the total emission—was calculated to be ~1 within the measured spectral window, which is among the highest reported values for quantum emitters.[9, 26] For second-order autocorrelation measurements, after being normalized by the Swabian Time Tagger software, the data were fitted with a three-level theoretical function:

$$g^{(2)}(\tau) = 1 - Ae^{\left(\frac{-|\tau|}{\tau_1}\right)} + Be^{\left(\frac{-|\tau|}{\tau_2}\right)}$$

Where $\tau_1$ and $\tau_2$ are the lifetimes at the excited and metastable state, while A and B denote the amplitude of antibunching and bunching behavior.[21, 27] If the bunching coefficient B is negligible, the given function reduces to the two-level form:

$$g^{(2)}(\tau) = 1 - Ae^{\left(\frac{-|\tau|}{\tau_0}\right)}$$

where $\tau_0$ is the excited-state lifetime.[28] To reliably extract lifetimes and other related coefficients, we selected the wide measurement range of approximately -1000 ns to 1000 ns, as shown in the middle panels of **Figure 4a-d**. The zoomed-in autocorrelation curves around the antibunching dip are shown in **Figure S6a-d** – **Supporting Information 5**, where clear antibunching features with the normalized coincidence at zero delay time $g^{(2)}(0)$ below 0.5 without any background correction are observed, confirming their single-photon nature. Additional examples of emitters with $g^{(2)}$ measurements are provided in **Figure S7 – Supporting Information 5**. Apart from single photon emitters (Emitter 1-4 in **Figure 4** and Emitter 5-6 in **Figure S7**), some cluster-like emitters with $g^{(2)}(0) > 0.5$ were also identified, exemplified by Emitter 7 and 8, **Figure S7**. This observation suggests the presence of both isolated single and clustered emission sites, which was also observed in emitters generated by other global or deterministic approaches.[18, 19, 23] Details of fitting parameters are displayed in **Table S3 – Supporting Information 5**. For the investigated emitters, the extracted $\tau_1$ was found to be from 0.697 ns to 5.271 ns, while the values for $\tau_2$ are in the range of 212.03 ns to 3012.37 ns, similar to previous reports. Several emitters have longer excited-state lifetimes (i.e., Emitter 1, 2, 5, 7, and 8), which may be related to their narrower linewidths. A similar correlation was also observed in another study.[21] Furthermore, several emitters with clear bunching features around the antibunching dip (Emitter 1, 5, 8) also exhibit higher bunching amplitudes B than the remaining ones.

Additionally, emission intensities versus excitation power of these emitters were also measured, and the acquired data were fitted to the function:

$$I = I_{sat} \times \frac{P}{P_{sat} + P}$$

Herein, $P_{sat}$ is the excitation power while $I_{sat}$ is the emission rate at saturation.[29] For Emitters 1 and 2, the saturation count rates are $9.92\times10^5$ and $5.89\times10^5$ count $s^{-1}$ at laser power of 0.246 and 0.905 mW, respectively, comparable to other reports.[14, 30] Interestingly, we found that Emitter 3 and Emitter 4 exhibited outstanding brightness, estimated at $4.47\times10^6$ and $12.25\times10^6$ count $s^{-1}$ at excitation powers of 1.542 and 4.965 mW, among the highest reported values, even without any photonic structures.[31]

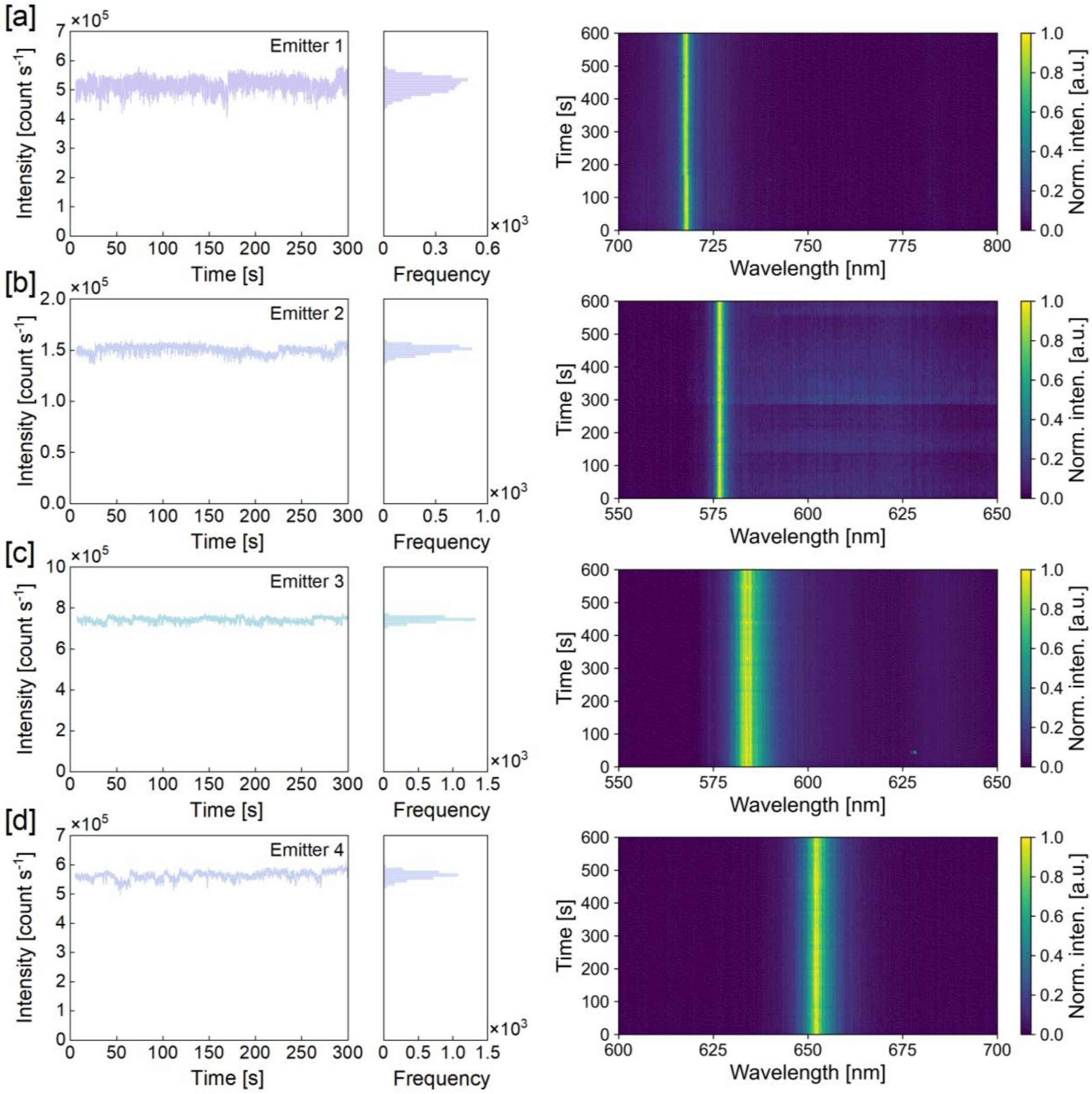

**Figure 5 |** Photostability measurements of hBN quantum emitters in overlap areas. Photon counts stability recorded over 300 seconds (left panels) and time traces of PL spectra of quantum emitter over 600 seconds (right panels) of (a) Emitter 1, (b) Emitter 2, (c) Emitter 3, and (d) Emitter 4.

The stability of Emitter 1-4 was also considered. As shown in the left graphs of **Figure 5a-d**, all emitters exhibit stable fluorescence emission over a 300-second recording period, with no blinking or bleaching. Their excellent stability is further supported by the spectral stability test, shown in the right panels of **Figure 5a-d**, where no significant spectral diffusion or blinking is detected over 10 minutes (600 seconds) of continuous measurement, comparing favorably with previous studies.[15, 18] It should be noted that the bright feature near the ZPL line of Emitter 2 can be ascribed to multiple phonon modes, as previously observed in its PL spectrum. Overall, these results demonstrate that the overlap approach effectively enhances emitter density without any evident degradation in optical performance.

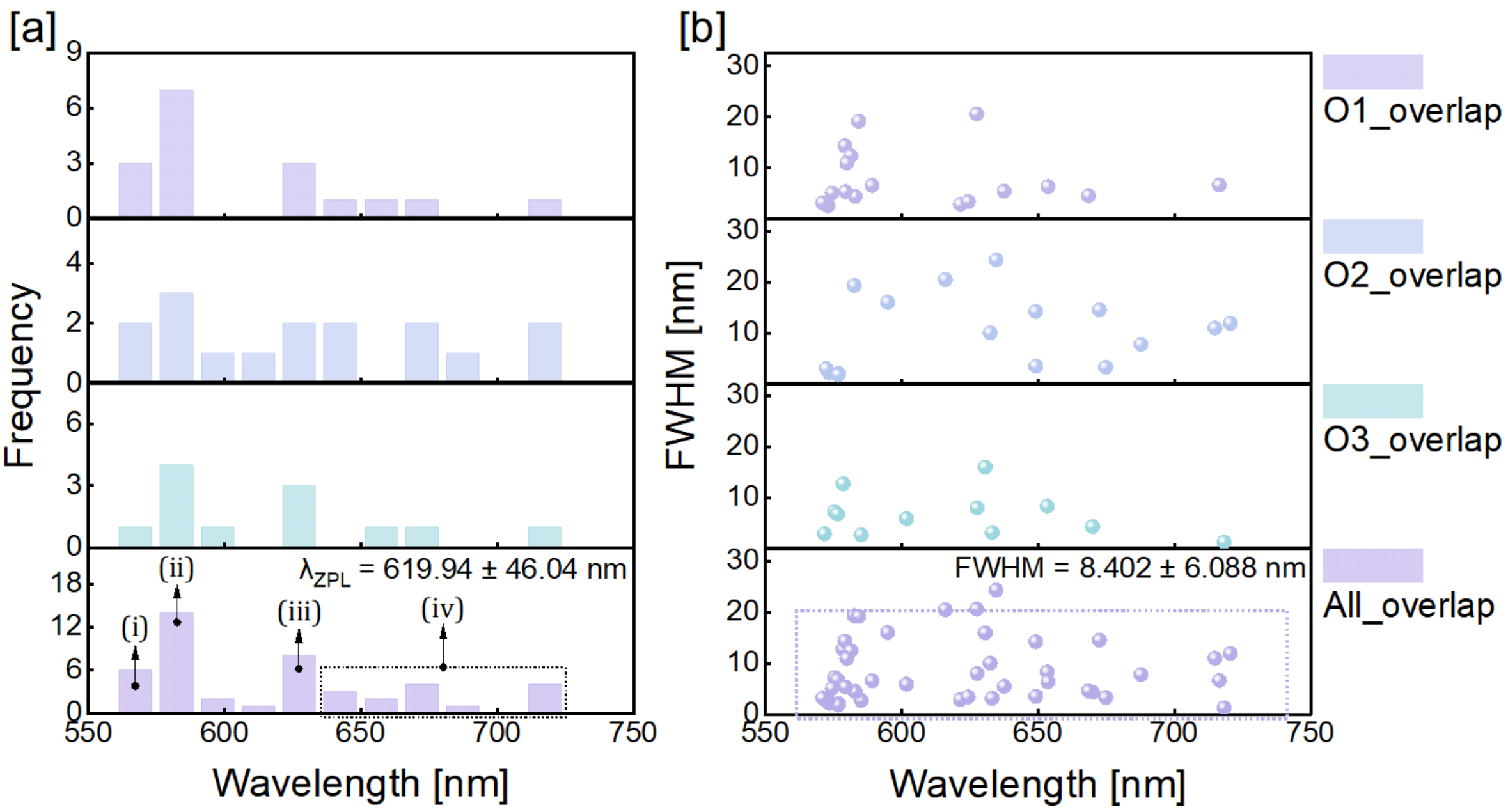

**Figure 6** | Statistical analysis of photoluminescence acquired from the emitters in the overlap regions. Histogram of (a) Zero-phonon line (ZPL) with the bin size of 10 nm and (b) Full width at half maximum (FWHM) with the bin size of 2 nm for 45 emitters found in the overlap area of samples O1-O3

We discuss the possible origin of the emitters found in the capped areas of samples O1-O3. As the exfoliation and subsequent transfer process may leave organic molecules (i.e., from tape), which have been previously identified as possible sources of single-photon emission,[32] we employed a high-temperature annealing in an oxygen atmosphere to promote their removal. To evaluate whether oxygen can access the overlap (capped) regions during annealing, we estimated the oxygen diffusion length in hBN flakes:

$$L = \sqrt{D \times t}$$

where $D$ is the diffusion coefficient, and $t$ is the diffusion time.[22] The total diffusion length during the ramp and hold stages, $L_{total}$, was estimated to be ~ 192 nm (cf. **Supporting Information 6** for details), which exceeds the thickness of the investigated flakes (in samples O1-O3). It is worth noting that the effective penetration depth may be longer than 192 nm, as oxygen can also diffuse through surface imperfections, edges, or interfacial gaps of flakes, suggesting that the organic residues were likely substantially removed. Therefore, we tentatively attributed the origin of the observed emitters in our study primarily to defects within hBN crystals. Additionally, the distribution of zero-phonon line wavelengths may provide further information about the possible defects. As shown in the histogram of the emitters' ZPL (**Figure 6a**), the emitters in the overlap areas of all samples show a wide emission range from 560 to 725 nm and the mean value of ZPL wavelength of all samples is 619.94 ± 46.04 nm. We noted that, the ZPL distributions of all samples shared similar features, suggesting the possible presence of several defect configurations. Based on the recurring ZPL distribution, we grouped into four spectral classes, denoted as (i), (ii), (iii), and (iv), and the possible origin of defects was discussed via comparison with previous studies.

The first class (i), accounting for 13.33% of the total number of emitters, may be attributed to the defect $V_NC_B$ or $C_BN_BV_N$; however, $V_NC_B$ is preferred as $C_BN_BV_N$ has been predicted to undergo a large structural change under excitation and to exhibit a high Huang-Rhys factor.[23, 33] The second emission class (ii), which dominates the emitter population with 31.11%, may originate from carbon trimers, particularly $C_BC_NC_N$ (or $C_2C_N$), initially introduced by *Jara et al*,[34] and then further studied by *Li* and colleagues.[35] Recent experiments also associated emitters in this configuration with emission lines around 565-590 nm.[19] In addition, recent studies also suggested that carbon complexes with different atomic arrangements ($C_BC_NC_BC_N$) are also potential candidates for classes (i) and (ii), with the ZPL centered around 562 nm and 573 nm.[17]

Class (iii), in the range of 620-635 nm, may be related to the intrinsic defects $N_BV_N$.[9] The final class (iv), with a broad emission wavelength distribution (~635 nm to beyond 700 nm), can be tentatively attributed to carbon- or oxygen-related configurations. The incorporation of oxygen into hBN crystals may result from high-temperature oxygen annealing, while previous research has also shown that it can be promoted by oxygen plasma treatment.[22, 27] Although some potential defect structures, such as $C_BC_NC_BC_N$, $O_BO_BV_N$, $V_BO_2$, have been proposed,[17, 27, 33] discrepancies between theoretical calculations and experiments keep these defect assignments under debate. Nevertheless, from an experimental perspective, emitters with ZPLs beyond 700 nm are often associated with oxygen-related defects.[22] Summary of the possible defects can be viewed in **Table S4** – **Supporting Information SI7**. In addition, statistics on the ZPLs of emitters from non-overlap areas also reveal similar information, with most emitters ascribed to C- and O-related defects (**Figure S5** – **Supporting Information SI7**). This suggests that our overlap geometry primarily enhances the probability of formation or preservation of optically active centers, rather than generating an entirely different defect configuration.

The estimated FWHM values of the emitters in the capped areas are presented for each sample and collectively for all samples, as displayed in **Figure 6b**. Most emitters exhibit linewidths below 20 nm, with several outstanding linewidths below 5 nm. Such observation across all samples further supports the reproducibility of our approach. Based on the emitters' characterization above, we believe that emitters generated via this facile transfer method are potential candidates for various photonics and sensing applications.

**Table 1 |**. Outcomes comparison between our and previous studies

| Materials | Treatment | ZPL | FWHM | Density | Brightness | Spectral stability | Ref |
|---|---|---|---|---|---|---|---|
| hBN flakes | Electron beam irradiation without post treatment | 575 | 19.56 ± 4.44 | - | $46.88\times10^3$ cps at 114 μW | - | 17 |
| hBN† flakes | Focus He ion beam and $O_2$ annealing (1050 ℃ for 1 hour) | 500-700 | - | - | $3.3\times10^5$ cps at 2.5 mW | 10 min | 16 |
| hBN flakes | Heavy ion irradiation and Ar annealing (850 ℃ for 1 hour) | 520-640 | - | 0.0213 | $4.3\times10^5$ cps at 3.5 mW | 100 s | 15 |
| hBN† flakes | Local stress without post treatment | 560-668 | 39.9 | - | - | 50 s | 18 |
| hBN† flakes | Nanoindentaiton and Ar annealing (1000 ℃ for 30 min) | 540-740 | 10-20 | - | - | - | 19 |
| hBN† flakes | Ar plasma and air annealing (850 ℃ for 30 min) | 550-600 | 17 | 0.045 | $25\times10^3$ cps at 700 μW | 160 s | 36 |
| hBN† flakes | $O_2$ plasma and $N_2$ annealing (850 ℃ for 30 min) | 560-650 | - | 0.028 | - | - | 23 |
| hBN† flakes | $O_2$ annealing (1100 ℃ for 4 hours) | 575-820 | 1.6-25 | - | $9.75\times10^5$ cps at 3.4 mW | 60 s | 22 |
| hBN flakes | Overlap and O2 annealing (1000 ℃ for 1 hour) | 560-725 | 8.4 ± 6.1 | 0.078 | $12.25\times10^6$ cps at 4.965 mW | 10 min | This study |

***Note**: All hBN flakes listed in this table were mechanically exfoliated from bulk crystals. The symbol † indicates commercially sourced hBN, and the remaining samples were grown by the respective authors. The table summarizes the best outcomes reported in each study, and the 'Spectral stability' column indicates the duration of the measurement.

***Unit**: ZPL and FWHM: nm, Density: $\mu m^{-2}$, cps: count per second (count $s^{-1}$)

**Table 1** compares our best results with those reported in previous studies. As in earlier reports, the emitters generated in our work exhibit visible-range emission, mainly attributed to intrinsic, carbon-related, or oxygen-related defects, while their FWHM values remain competitive. In addition, although our approach relies on ultrahigh-purity hBN, the achieved emitter density is comparable to that reported in studies using commercial hBN sources, which are generally expected to host various pre-existing defects. Furthermore, in addition to the significant enhancement in emitter density, our overlap approach does not introduce any apparent trade-off in emitter quality, as the brightness and spectral stability of the obtained emitters are comparable to, or even better than, those reported in some previous studies. This indicates that our simple, cost-effective transfer technique can be further employed in generating emitters with high density and spectral quality for quantum applications.

## 3. Conclusion

In summary, we proposed a facile all-dry transfer technique to form an overlap geometry from two hBN flakes. Despite its simplicity, this technique enables the generation of quantum emitters with enhanced density in transfer-defined overlap regions. The emitter density was significantly improved, reaching up to 15.8 times that of the non-overlap regions. Notably, several of these emitters exhibited outstanding SBR, SNR, and DW factor, and representative emitters showed competitive brightness and spectral stability compared with previous reports. The emitter density enhancement cannot be explained solely by the mere presence of an additional flake, but is likely associated with the capping effect introduced by the overlap geometry. Furthermore, statistical analysis of spectral quality metrics indicates that the transfer process does not noticeably compromise the emitter spectral quality compared with the control areas. Our approach sheds light on the generation of high-density, high-brightness, and highly spectrally pure quantum emitters for photonics and sensing applications.

## 4. Experimental details

### 4.1. Samples fabrication

Hexagonal boron nitride bulk crystals were synthesized at 1500 °C under a high pressure of 4.5 GPa to obtain a low concentration of carbon and oxygen impurities below < $10^{18}$ $cm^{-3}$.[37] The flakes were exfoliated from these ultrahigh-purity bulk crystals using adhesive tape and subsequently transferred onto silicon substrates with a 285-nm thermal oxide layer on top ($Si/SiO_2$). Please note that prior to exfoliation, the substrates were pre-cleaned using a standard sonication process in acetone and isopropyl alcohol (IPA) for 5 minutes each, followed by an air-gun blow-off. The hBN samples were consequently localized on an x-y-z stage (stage A, **Figure S1a**) for subsequent transfer.

The deterministic transfer process was inspired by a previous study.[38] Initially, the exfoliated hBN flakes on the same adhesive tape were transferred to a polydimethylsiloxane (PDMS) elastic stamp mounted on a glass slide to ensure that the top and bottom flakes originate from the same bulk crystal. This glass slide was then mounted onto another x-y-z stage (denoted as stage B, **Figure S1a**). Targeted flakes on the $Si/SiO_2$ substrate and on the PDMS stamp were pre-selected and aligned relative to each other under optical microscope guidance. The glass slide was then slowly lowered until the flake on PDMS approached the flake on $Si/SiO_2$ substrate. After 5 minutes of contact, the glass slide was gently lifted, leaving the overlap hBN flakes on the substrates. The transfer process was performed using the setup shown in **Figure S1a** without applying any external heat.

Finally, all the samples were annealed at 1000 °C under an oxygen atmosphere (200 sccm) in a *Lindberg* tube furnace to activate and generate quantum emitters. During annealing, the samples were placed at the center of the tube, heated at a ramp rate of ~8 °C/min, held at 1000 °C for 1 hour, and then naturally cooled to room temperature (~4 hours).

In our study, the entire experimental process was repeated independently using different original hBN crystals to assess the reproducibility of the transfer approach.

### 4.2. Optical measurements and material characterization

In the present study, the confocal maps, photoluminescence (PL) spectra, and second-order autocorrelation were obtained at room temperature using a lab-built confocal setup (the schematic illustration and details of the setup are presented in **Figure S1b**). Surface topography of the samples were measured using an atomic force microscope (AFM) images recorded by the system (Nanosurf DriveAFM) in tapping mode, using Tap150-G cantilever with a resonance frequency of 152 kHz. The raw images were subsequently processed in Gwyddion to extract the flakes thicknesses.

## ASSOCIATED CONTENT

### Data Availability Statement

The experimental data acquired during this study is available from the corresponding authors upon reasonable request.

### Supporting Information

The Supporting Information includes: 1. Schematic illustration of the lab-built all-dry transfer and confocal scanning setups, 2. Confocal scanning and PL spectra of the resulting samples, 3. Calculation of areal and thickness-normalized emitter density, 4. Calculation of spectral quality metrics: signal-to-background and signal-to-noise ratio (SBR and SNR), and Debye-Waller (DW) factor, 5. Photon-correlation characterization and additional examples of emitters, 6. Calculation of oxygen diffusion length into hBN flakes, 7. Possible origins of emitters.

## AUTHOR INFORMATION

Corresponding Author

† Corresponding author: trongtoan.tran@uts.edu.au

Author Contributions

**N.M.N**: Conceptualization, Methodology, Data curation, Formal analysis, Writing – original draft. **T.V.D**, **M.S.H**, **A.E, H.N.D.H** and **T.N.A.M**: Methodology, Data curation, Writing – original draft. **D.A.N** and **Y.C**: Formal Analysis, Writing – review & editing. **K.W** and **T.T**: Methodology, Writing – review & editing. **M.G.R**: Data curation, Writing – review & editing. **C.C**, **X.X**, and **T.D**: Formal analysis, Writing – review & editing. **T.T.T**: Conceptualization, Formal analysis, Funding acquisition, Project administration, Supervision, Writing – review & editing.

**Funding Sources**

T. T. T acknowledges the financial support from the Australian Research Council (DE220100487, DP240103127). M.G.R acknowledges the financial support from the Australian Research Council (DE240100507). C.C acknowledges the financial support from the Australian Research Council (DE250100406). T. T. T. and T. D. thank the Queensland Department of Environment, Science, and Innovation for their financial support (Q2032010). This project was funded by the Queensland Government through the Department of Environment, Tourism, Science and Innovation's (DETSI) Quantum 2032 Challenge Program. The program aims to accelerate the development of quantum-based innovations in sportstech and related fields, foster collaboration between Queensland's quantum research sector and industry, and showcase the state's quantum expertise on the global stage during the Brisbane 2032 Olympic and Paralympic Games, contributing to the lasting legacy of the Games. K.W. and T.T. acknowledge support from the CREST (JPMJCR24A5), JST and World Premier International Research Center Initiative (WPI), MEXT, Japan. This research is supported by an Australian Government Research Training Program (RTP) Scholarship.

**Notes**

The authors declare no competing financial interest.

Supporting Information

# Facile hBN–hBN Interfacial Overlap Engineering for Enhanced Quantum Emitter Formation

Nhat Minh Nguyen [1], Trung Vuong Doan [1], Md Shakhawath Hossain [1], Akila Elangasinghe [2], Duc Anh Ngo [1], Ha Ngoc Duy Huynh [1], Thi Ngoc Anh Mai [1], Yongliang Chen [3], Kenji Watanabe [4], Takashi Taniguchi [5], Michael G. Ruppert [2], Chaohao Chen [1], Xiaoxue Xu [1], Toan Dinh [6,7], and Toan Trong Tran [1,†]

[1] School of Electrical, Mechanical and Biomedical Engineering, University of Technology Sydney, Ultimo, NSW, 2007, Australia

[2] Centre for Audio, Acoustics and Vibration, University of Technology Sydney, Ultimo, NSW 2007, Australia

[3] Department of Physics, The University of Hong Kong, Pokfulam, Hong Kong, China

[4] Research Center for Electronic and Optical Materials, National Institute for Materials Science, 1-1 Namiki, Tsukuba 305-0044, Japan

[5] Research Center for Materials Nanoarchitectonics, National Institute for Materials Science, 1-1 Namiki, Tsukuba 305-0044, Japan

[6] School of Engineering, University of Southern Queensland, Toowoomba, Queensland 4350, Australia

[7] Center for Future Materials, University of Southern Queensland, Toowoomba, Queensland 4350, Australia

† Corresponding author: Toan Trong Tran | trongtoan.tran@uts.edu.au

## 1. Schematic illustration of the lab-built all-dry transfer and confocal scanning setups

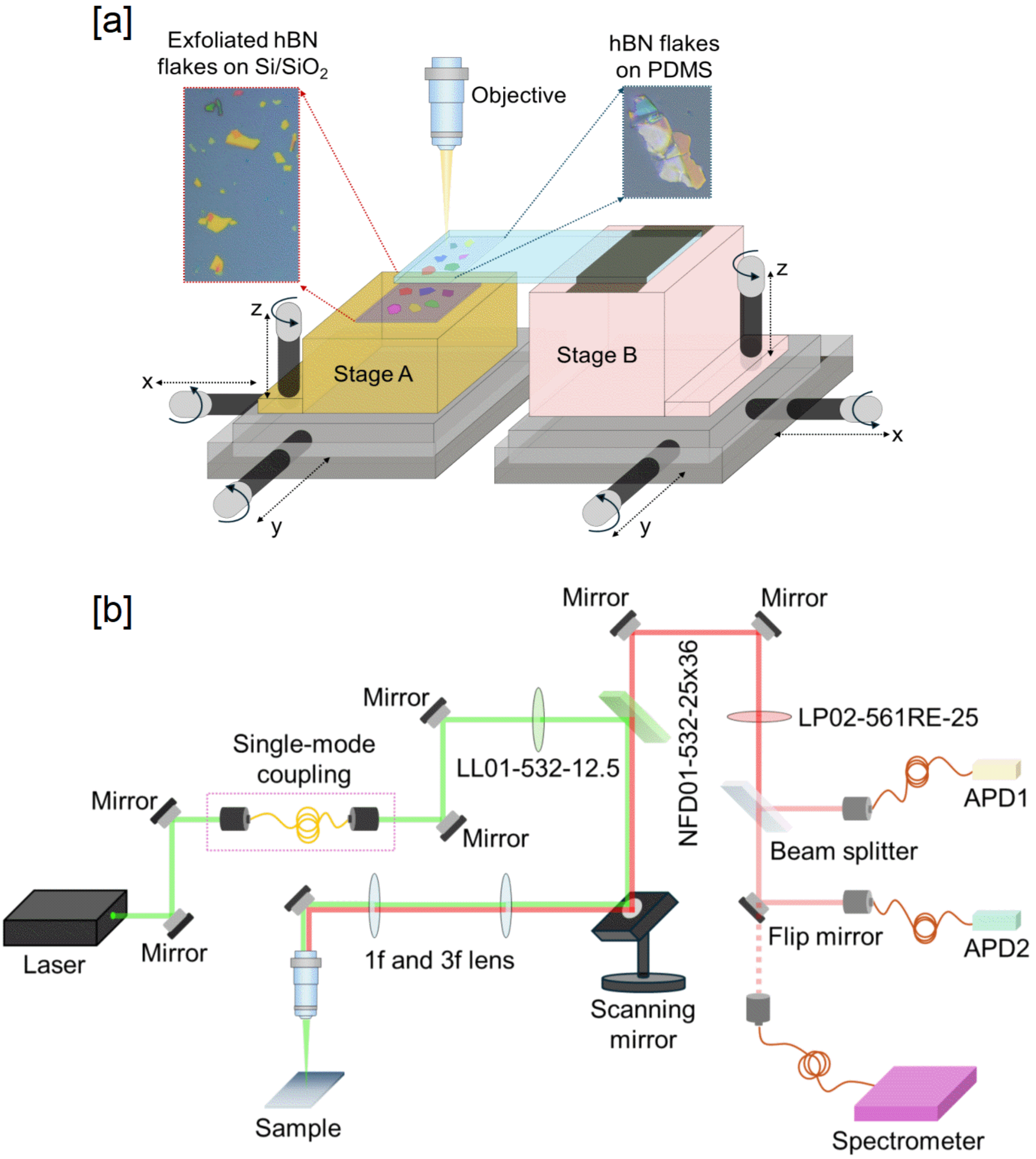

**Figure S1** | Schematic illustration of the lab-build **[a]** All-dry transfer system and **[b]** Confocal scanning set up. The yellow and orange cables represent the single and multi-mode fibers, respectively. The long pass filter in the collection pathway (LP02-561RE-25) was tilted 30º to achieve wider spectra.

- Transfer setup. **Figure S1a** presents our lab-built all-dry transfer setup used in this study. Here, two x-, y-, and z-axis stages are employed to mount the sample and the glass carrying the PDMS stamp. The transfer was performed by adjusting the x-, y-, and z- knobs in both stages under the guidance of an optical microscope.

- Confocal scanning setup. **Figure S1b** shows the schematic of our lab-built confocal scanning system. In detail, a green laser from a continuous-wave (532 nm, Cobolt Samba) excitation source is coupled into a single-mode fiber to maintain the spatial mode quality, then directed and focused onto the sample via a 4f lens system and a 100× objective (NA=0.7; Thorlabs, MY100X-806). In the collection pathway, a long-pass filter (Semrock, LP02-561RE-25) tilted by $30^{o}$ is placed to eliminate any residual laser leakage. During measurement, a scanning mirror (Newport SFM-CD300B) steers the laser beam across the sample while the incoming photons are collected by an avalanche photodiode (APD), and a spectrometer acquires the PL spectra. A Hanbury Brown and Twiss configuration with two APDs is employed to measure the second-order autocorrelation function [$g^{(2)}(\tau)$], and the data is directly normalized by the Swabian Time Tagger software. A flip mirror is utilized to switch between collecting PL spectra or [$g^{(2)}(\tau)$] curves, and the whole setup is connected to a computer for analysis. All the PL spectra were collected under identical parameters (i.e., room temperature, similar laser power, and acquisition time). A bright spot on the confocal map was considered an emitter when its PL spectrum showed an SNR $\geq$ 3 (the signal detection threshold).

## 2. Confocal scanning and PL spectra of the resulting samples

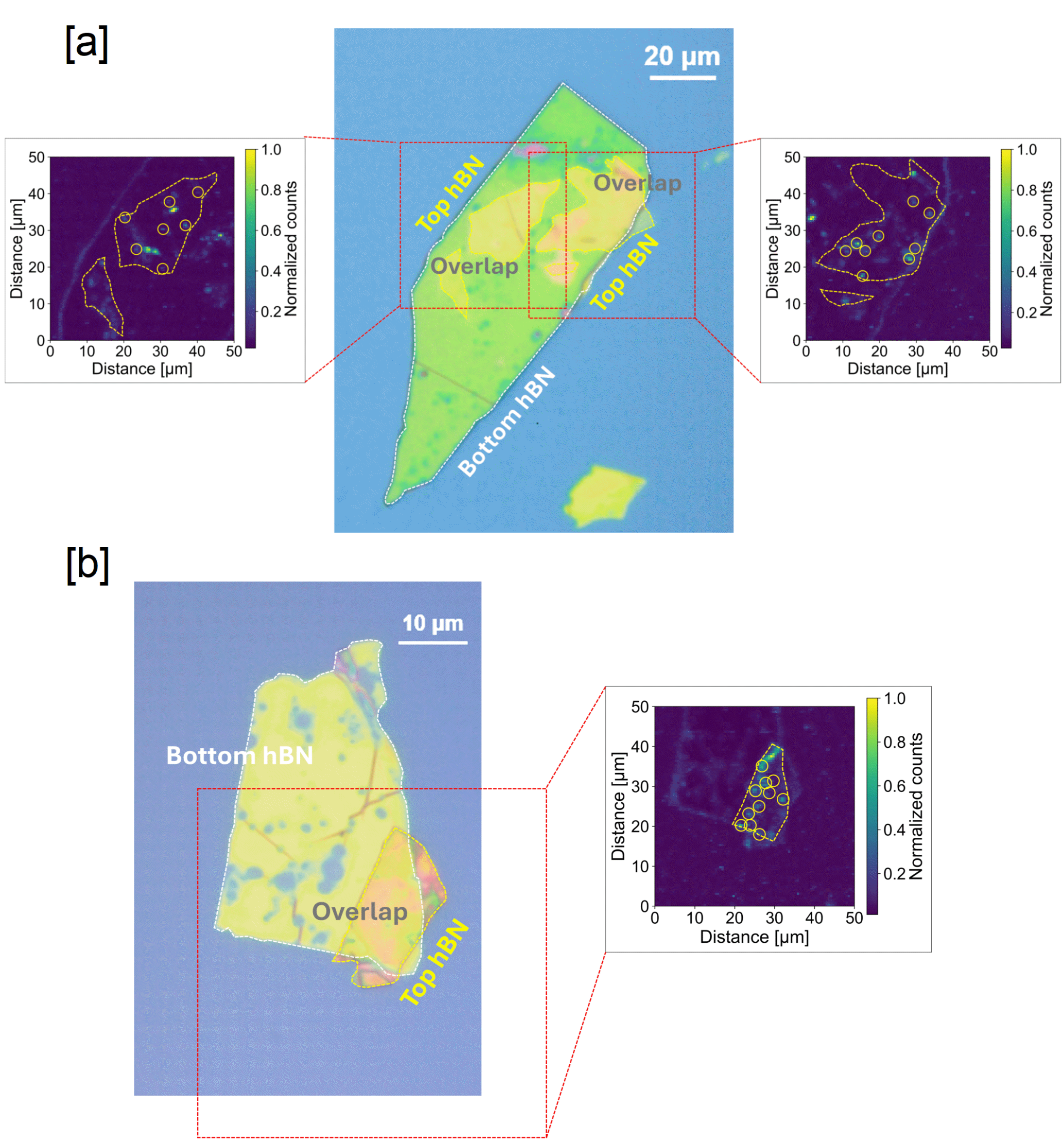

**Figure S2** | PL spectra of representative emitters from overlap regions in samples **[a]** O2, **[b]** O3 (images of sample O1 are presented in **Figure 2a** – main text). In the confocal maps, the yellow circles indicate the position where PL spectra were collected.

- **Figure S2a-b** shows the optical images and the confocal scanning maps of the corresponding overlap areas of samples O2 and O3. The dashed outlines indicate the regions where the identified emitters were counted for overlap.

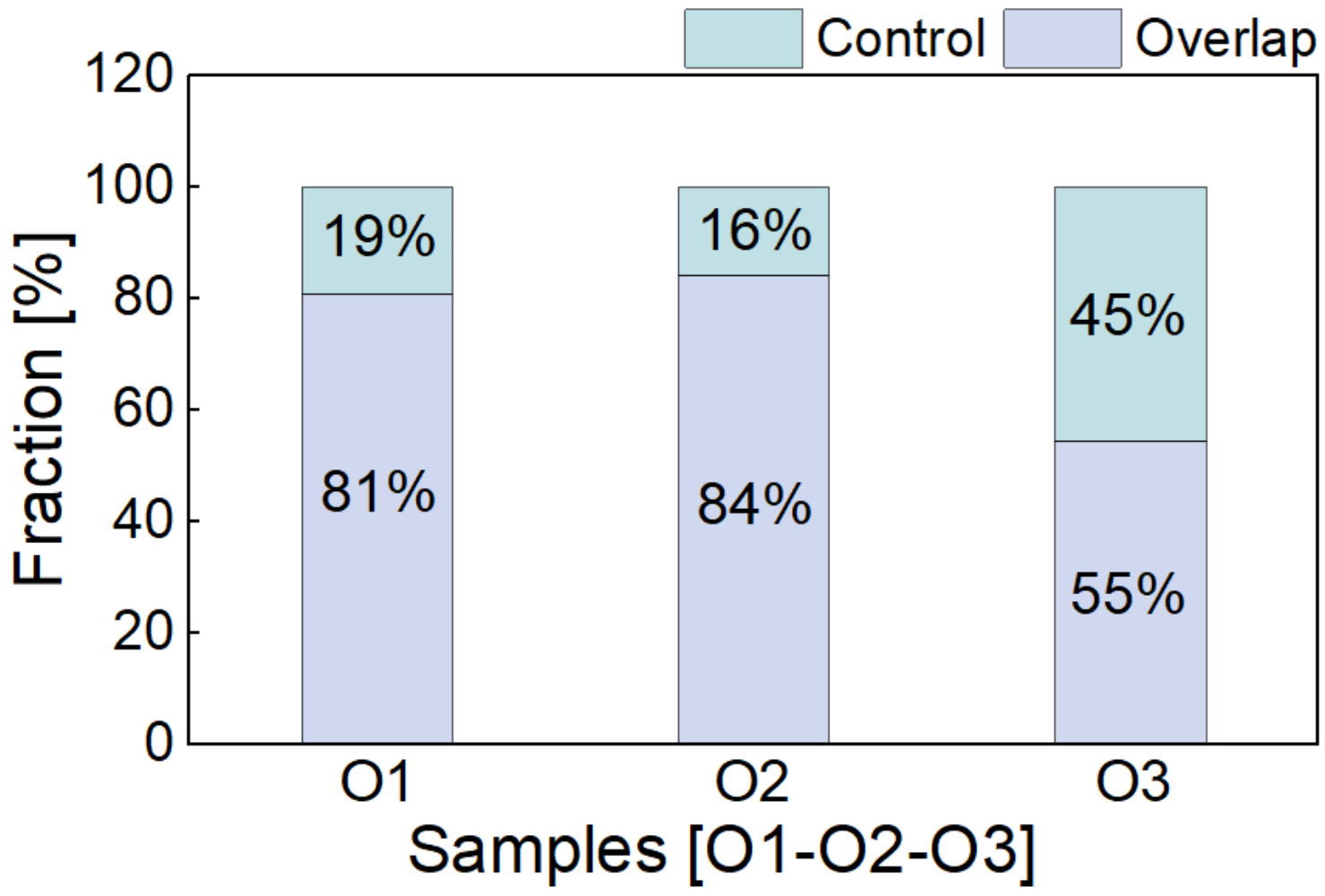

**Figure S3** | Fraction of emitters found in overlap and control regions in samples O1, O2 and O3

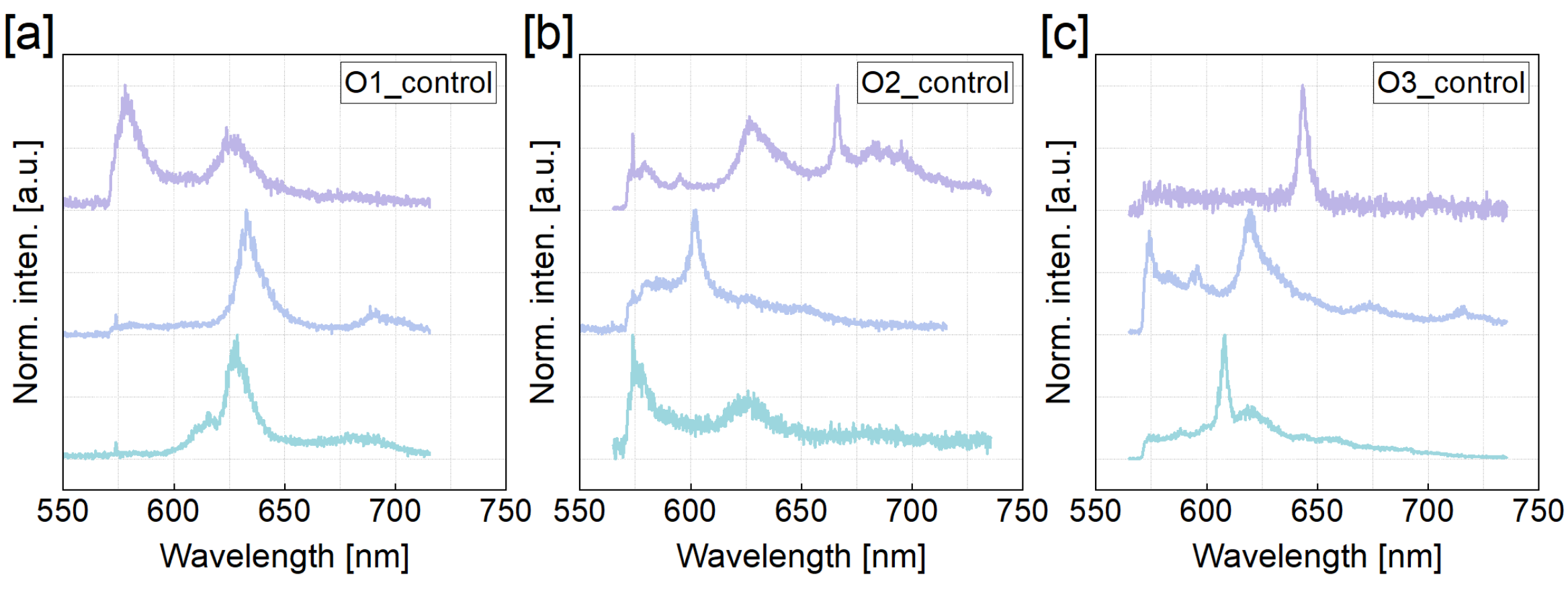

**Figure S4** | PL spectra of representative emitters in control regions of samples (a) O1, (b) O2 and (c) O3

- **Figure S4a-c** presents the PL spectra of representative emitters from control (non-overlap) regions. In general, although we were able to find emitters in these regions, several of them exhibit the multi-peak and less well-resolved profiles, suggesting that some uncapped emission centers may be more sensitive to the high-temperature oxygen annealing process.

## 3. Calculation of areal and thickness-normalized emitter density

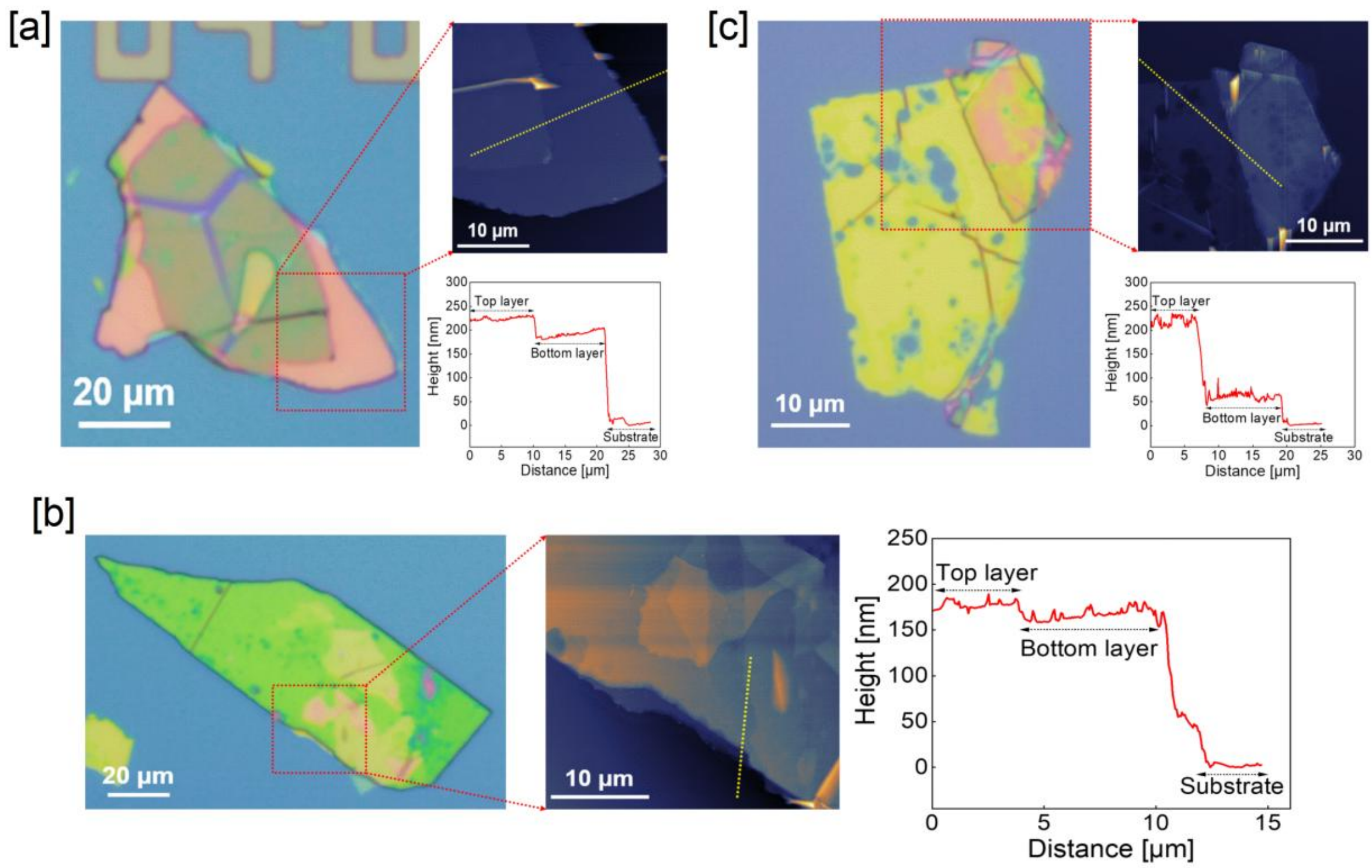

**Figure S5** | Optical images, AFM images, and the extracted thickness of the flakes of samples (a) O1, (b) O2, and (c) O3. The yellow dashed lines indicate the direction in which the height profiles were acquired.

- Optical images of the three samples after annealing are shown in **Figure S5a-c**. It can be seen that annealing at high temperature in oxygen may induce some etching, especially in thin flakes (obvious in the optical image of **Figure S5c**). This is consistent with the study by Li et al., which reported that hBN flakes may oxidize in air when the temperature exceeds 850 °C.[1]

- Areal emitter density. Initially, the non-overlap and overlap regions were obtained using ImageJ, with the optical images with scale bars shown in **Figure S5a-c** as input. The area of each region was calculated 10 times and presented as the mean ± SD (standard deviation) to estimate the uncertainty. Note that the visibly etched area after annealing was excluded from the calculations. Then, the emitter's density was acquired by:

$$\rho_a = \frac{N}{A}$$

Where N is the number of emitters, and A is the area in question.[2] Detailed results of the calculation are presented in **Table S1**.

**Table S1** | Calculated area and emitter density for non-overlap and overlap regions of samples O1, O2 and O3

| Sample | Area [μm$^2$] | | Areal emitter density [μm$^{-2}$] | |
|---|---|---|---|---|
| | **Control** | **Overlap** | **Control** | **Overlap** |
| O1 | 908.87 ± 7.89 | 1537.99 ± 6.64 | 0.0044 ± 0.0022 | 0.0111 ± 0.0027 |
| O2 | 3092.63 ± 12.85 | 1010.85 ± 6.00 | 0.0010 ± 0.0006 | 0.0158 ± 0.0040 |
| O3 | 853.47 ± 4.89 | 153.95 ± 0.969 | 0.0124 ± 0.0039 | 0.0780 ± 0.0225 |

Apart from the standard deviation of the areas, the Poisson uncertainty associated with the emitter counts, taken as $\sqrt{N}$ was also included in the density uncertainty.[3]

- Thickness-normalized emitter density. We also calculated the thickness-normalized emitter density to evaluate whether the increased emitter density simply originates from the increased hBN thickness in the overlap region. From the AFM images (**Figure S5a-c**), the flake thicknesses were measured from 10 height profiles extracted from the scanned area using Gwyddion software and presented as mean ± SD (standard deviation). Then, the thickness-normalized density is calculated by:

$$\rho_t = \frac{N}{A \times t} = \frac{\rho_a}{t}$$

where t is the hBN effective thickness, i.e., the thickness of the bottom hBN layer for the control regions and the sum of the top and bottom layers thicknesses for the overlap regions. The results are shown in **Table S2**.

**Table S2** | Average thicknesses of top and bottom layers and the calculated thickness-normalized emitter density for control and overlap regions in samples O1, O2 and O3

| **Sample** | **Average thickness [nm]** | | **Thickness-normalized emitter density [µm$^{-3}$]** | |
|---|---|---|---|---|
| | **Top layer** | **Bottom layer** | **Control** | **Overlap** |
| O1 | 31.73 ± 3.83 | 192.06 ± 10.54 | 0.0229 ± 0.0115 | 0.0494 ± 0.0122 |
| O2 | 7.37 ± 1.25 | 157.93 ± 5.76 | 0.0061 ± 0.0036 | 0.0958 ± 0.0242 |
| O3 | 163.17 ± 13.24 | 63.59 ± 4.33 | 0.1957 ± 0.0633 | 0.3438 ± 0.1015 |

In all three samples, the overlap regions still exhibit higher thickness-normalized emitter density than the corresponding control regions, suggesting that additional factors beyond the increased investigated area and effective hBN thickness may contribute to the improved emitter density, as discussed in the main text.

**4. Calculation of spectral quality metrics: signal-to-background and signal-to-noise ratio (SBR and SNR), and Debye-Waller (DW) factor**

To demonstrate that our overlap approach does not introduce a significant trade-off in emitter quality, we calculated and compared typical spectral quality metrics of emitters found in both regions.

- SBR and SNR. For SBR and SNR, we randomly selected 100 points from the baseline of a PL spectrum (not including the PSB). The values of these points are assigned as: $Y_1, Y_2, \dots Y_i, \dots Y_{100}$

For consistency, herein, we defined:

$\mathbf{I_{signal,total}}$: Total intensity within the ZPL spectral window (including signal from ZPL and the underlying background)

$\mathbf{I_{signal}}$: The signal from the ZPL only

$\mathbf{I_{background}}$: the signal from the background only (exclude the PSB)

*Signal-to-background ratio is a measure of how strong the signal is compared to the background, and can be calculated following the study of Zhai et al [4]

$$\mathrm{SBR} = \frac{I_{signal}}{I_{background}} = \frac{I_{signal,total} - I_{background}}{I_{background}}$$

Where $I_{background} = \overline{Y} = \frac{\sum_{i=1}^{N} Y_i}{N}$ is the average background of 100 random points ($N = 100$) from the background as mentioned above. Note that $I_{signal,total} = 1$ because all the spectrum was normalized.

*SNR reflects how reliably the signal can be distinguished from the noise. This parameter plays a crucial role in analytical applications, and is expressed by:

$$\mathrm{SNR} = \frac{\mathrm{I_{signal}}}{\mathrm{RMS_{Noise}}} = \frac{\mathrm{I_{signal,tot}} - \mathrm{I_{background}}}{\mathrm{RMS_{Noise}}}$$

With RMS, the noise is the quantification of the noise level of the measurement,[5] and is calculated by:

$$\mathrm{RMS_{Noise}} = \sqrt{\frac{\sum_{i=1}^{N}(\overline{Y} - Y_i)^2}{N}}$$

Where $\overline{Y}$ is the average value of 100 random baseline points ($N = 100$). In this study, to ensure accuracy, each parameter was calculated 10 times, and the average value was reported.

- Debye-Waller factor. DW factor is the ratio between the photons contributing to the ZPL signal and the total number of emitted photons. This parameter is used to evaluate if an emitter exhibits the strong ZPL and weak PSBs,[6] and can be acquired using the formula:

$$\mathrm{DW} = \frac{\int \mathrm{I_{ZPL}}\, d\lambda}{\int \mathrm{I_{ZPL}}\, d\lambda + \int \mathrm{I_{PSB}}\, d\lambda}$$

As we have $I = \int_{\lambda_1}^{\lambda_2} I\, d\lambda$ is the area under the fitting curve in PL spectrum. Hence, DW factor is the ratio of the integrated area of ZPLs component to the total fitted emission area:

$$\mathrm{DW} = \frac{\Sigma \mathrm{A_{ZPLs}}}{\Sigma \mathrm{A_{ZPLs}} + \Sigma \mathrm{A_{PSBs}}}$$

In the present study, the background was accounted for using an offset parameter in the Lorentzian fitting model, and the integrated fitted areas correspond only to the contributions of the ZPLs and PSBs.

- Monte Carlo subsampling analysis. To account for the unequal number of emitters identified in the overlap and control regions, we applied a Monte Carlo random subsampling analysis.[7] For each subsampling trial (k), a subset ($S_k$) containing $N_{control} = 17$ emitters were randomly selected without replacement from 45 emitters in overlap regions, and the spectral quality metrics of these randomly selected emitters were calculated as:

$$\overline{M}_{overlap}^{(k)} = \frac{1}{N_{control}} \sum_{i \in S_k} M_{overlap,i}$$

The resulting values of $\overline{M}_{overlap}^{(k)}$ which corresponds to the parameters being considered (SBR, SNR, and DW factor) are used to construct the histogram for further comparison, as shown in the main text.

**5. Photon-correlation characterization and additional examples of emitters**

- Typical three-level model for quantum emitters:

$$g^{(2)}(\tau) = 1 - Ae^{\left(\frac{-|\tau|}{\tau_1}\right)} + Be^{\left(\frac{-|\tau|}{\tau_2}\right)}$$

$\tau_1$ and $\tau_2$ are the lifetimes of the excited and metastable or shelving states; A and B represent the amplitudes of the antibunching and bunching features.[8, 9] The shape of the autocorrelation function can be reflected by A and B, accordingly.

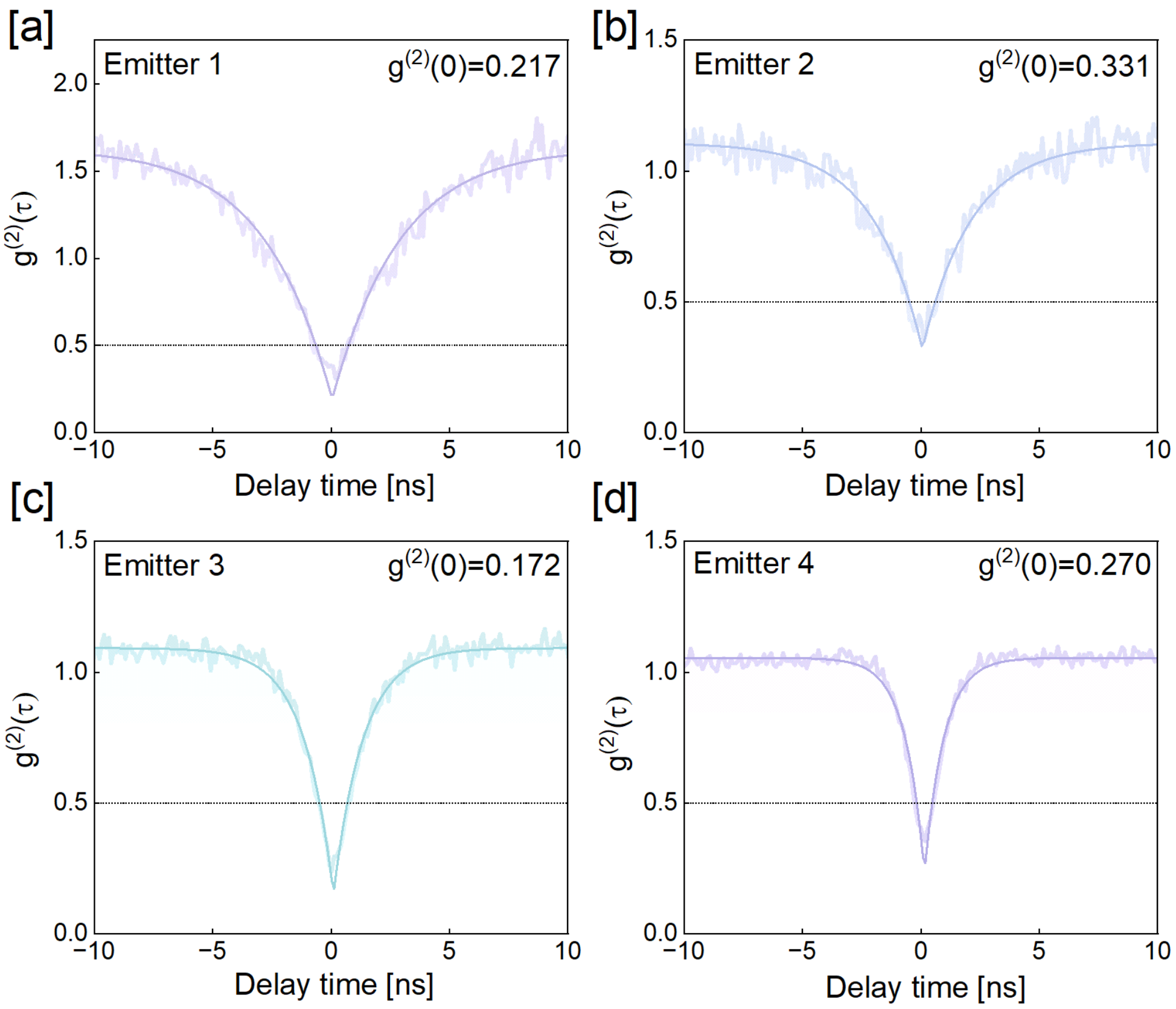

**Figure S6** | Zoomed-in auto correlation around the antibunching dip of (a) Emitter 1, (b) Emitter 2, (c) Emitter 3, and (d) Emitter 4

- **Figure S6a–d** shows magnified views of the autocorrelation measurement presented in the middle panels of **Figure 4a–d**. The antibunching dips below 0.5 at zero delay time are clearly observed in these magnified plots, indicating the single-emission behavior.

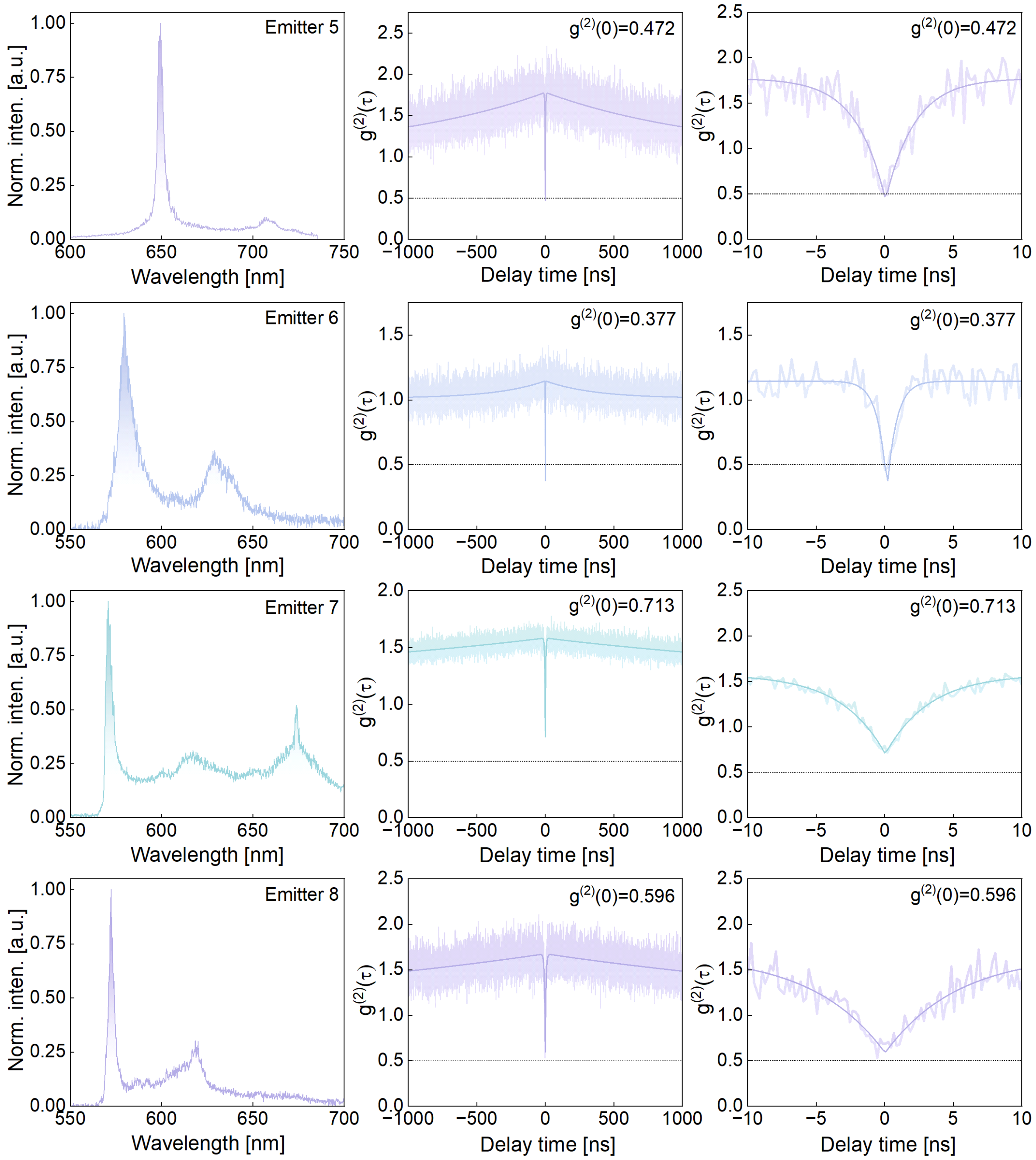

**Figure S7** | PL spectra (left panels), long-delay (middle panels) and zoom-in (right panels) auto-correlation measurements of some other emitters from samples O1-O3

- PL spectra and $g^{(2)}$ measurements of additional emitters are shown in **Figure S7**. In addition to some single emitters with $g^{(2)}(0)<0.5$, cluster-like emitters were also observed (Emitter 7-8). Such emitters exhibit multi-peak PL profiles with $g^{(2)}(0)>0.5$ and pronounced bunching behaviors.

**Table S3** | Fitting parameters from autocorrelation function for typical emitters 1-8

| Emitter | $\tau_1$ [ns] | $\tau_2$ [ns] | A | B |
|---|---|---|---|---|
| E1 | 2.905 | 231.235 | 1.470 | 0.564 |
| E2 | 2.116 | 212.028 | 0.790 | 0.062 |
| E3 | 1.258 | 331.006 | 0.944 | 0.090 |
| E4 | 0.827 | 1026.150 | 0.823 | 0.046 |
| E5 | 2.141 | 1227.695 | 1.346 | 0.750 |
| E6 | 0.697 | 352.269 | 0.823 | 0.134 |
| E7 | 3.230 | 2587.115 | 0.881 | 0.383 |
| E8 | 5.271 | 3012.368 | 1.090 | 0.676 |

- Fitting results in **Table S3** reveal that the lifetimes are in the range of 0.697 ns to 3.230 ns, in correlation with previous reports. For emitters with pronounced bunching features (around the anti-bunching dip), their B values are generally higher than those of others. It is worth noting that two emitters with a high value of B still satisfy $g^{(2)}(0)<0.5$ as they have a high value of A (anti-bunching coefficient).

## 6. Calculation of oxygen diffusion length into hBN flakes

- As the first step to discuss the possible origin of the generated emitters, we calculated the diffusion length of oxygen into hBN using the formula:

$$L = \sqrt{D \times t}$$

where D is the diffusion coefficient, depending on the nature of the material and the working temperature, and t is the time of diffusion.[10] We can find D at a specific temperature using the Arrhenius equation:

$$D = D_0 \exp\left(\frac{-E_A}{RT}\right)$$

Here, $E_A$ is the activation energy, R is the universal gas constant and T is the working temperature. In previous studies, the diffusion coefficients at 1450 °C and 1100 °C were calculated to be $2\times10^{-17}$ m$^2$/s and $9.4\times10^{-18}$ m$^2$/s, respectively.[10, 11] From here, $E_A$ and $D_0$ are found to be ~$4.24\times10^4$ J.mol$^{-1}$ and ~$3.86\times10^{-16}$ m$^2$/s.

*The diffusion length for the annealing process at 1000 °C for 1 hour:

$$L_{hold} = \sqrt{D_0 \exp\left(\frac{-E_A}{R \times T_{hold}}\right) \times t_{hold}}$$

$$= \sqrt{3.86 \times 10^{-16} \times \exp\left(\frac{-4.24 \times 10^4}{8.31 \times (1000 + 273.15)}\right) \times 3600}$$

$$= 1.59 \times 10^{-7} (\mathrm{m}) = 159\ (\mathrm{nm})$$

*However, since the ramping is relatively long (~2 hours), we also take the diffusion length of this stage into account. As the temperature changed continuously, we began with:

$$L_{ramp} = \sqrt{D \times t} \leftrightarrow {L_{ramp}}^2 = D \times t \leftrightarrow d({L_{ramp}}^2) = D(t) \times dt$$

$$\leftrightarrow \int d\left(L_{ramp}{}^{2}\right) = \int D(t) \times dt \leftrightarrow L_{ramp}{}^{2} = \int D(t) \times dt = \int D_0 \exp\left(\frac{-E_A}{RT_{ramp}}\right) \times dt_{ramp}$$

Assume that the ramp process was linear with a rate of $\beta = 8\left(°C\frac{}{min}\right) = \ 8\ \left(\frac{K}{min}\right) = \frac{8}{60}\ (K/s)$, we have:

$$T_{ramp} = T_0 + \beta \times t_{ramp} \leftrightarrow dt_{ramp} = \frac{dT_{ramp}}{\beta}$$

From here:

$$L_{ramp}{}^{2} = \int D_0 \exp\left(\frac{-E_A}{R \times T_{ramp}}\right) \times dt_{ramp} = \int D_0 \exp\left(\frac{-E_A}{R \times T_{ramp}}\right) \times \frac{dT_{ramp}}{\beta}$$

For the process from room temperature (298.15 K) to 1000 °C (1273.15 K), we have:

$$L_{ramp}{}^{2} = \int_{298.15}^{1273.15} D_0 \exp\left(\frac{-E_A}{R \times T_{ramp}}\right) \times \frac{dT_{ramp}}{\beta} = 1.15 \times 10^{-14} (m^2)$$

$$\leftrightarrow L_{ramp} = 107\ (nm)$$

*The total effective diffusion length was estimated by considering the contributions from the ramping and holding stages,[12] which were combined in quadrature as:

$$L_{total} = \sqrt{L_{ramp}{}^{2} + L_{hold}{}^{2}} = 192\ (nm)$$

*Please note that the practical diffusion length exceeds 192 nm, as the flakes inevitably contain surface imperfections, cracks, and lateral diffusion may also contribute. Since the total diffusion length exceeds the flake thicknesses, organic molecules are likely to be removed during annealing. Hence, we tentatively attribute the emitters in the current study to imperfections in the hBN crystal.

## 7. Possible origins of emitters

**Table S4** | Summary of the possible defect origins of emitters in the overlap region, compiled from the literature.

| No | Bin [nm] | Fraction [%] | Possible origin | Ref |
|---|---|---|---|---|
| 1 | 560 – 575 | 13.33 | $V_NC_B$, $C_BN_BV_N$, $C_BC_NC_BC_N$ | 13-15 |
| 2 | 575 – 590 | 31.11 | $C_BC_NC_N$ (or $C_2C_N$), $C_BC_NC_BC_N$ | 15, 16 |
| 3 | 590 – 605 | 4.44 | C-related defects (not specifically assigned yet) | - |
| 4 | 605 – 620 | 2.22 | | |
| 5 | 620 – 635 | 17.78 | $N_BV_N$ (intrinsic) | 17 |
| 6 | 635 – 650 | 6.67 | C- or O-related defects<br>Potential candidates: $C_BC_NC_BC_N$, $O_BO_BV_N$, $V_BO_2$<br>(not fully assigned yet) | 8, 13, 15 |
| 7 | 650 – 665 | 4.44 | | |
| 8 | 665 – 680 | 8.89 | | |
| 9 | 680 – 695 | 2.22 | | |
| 10 | 710 – 725 | 8.89 | | |

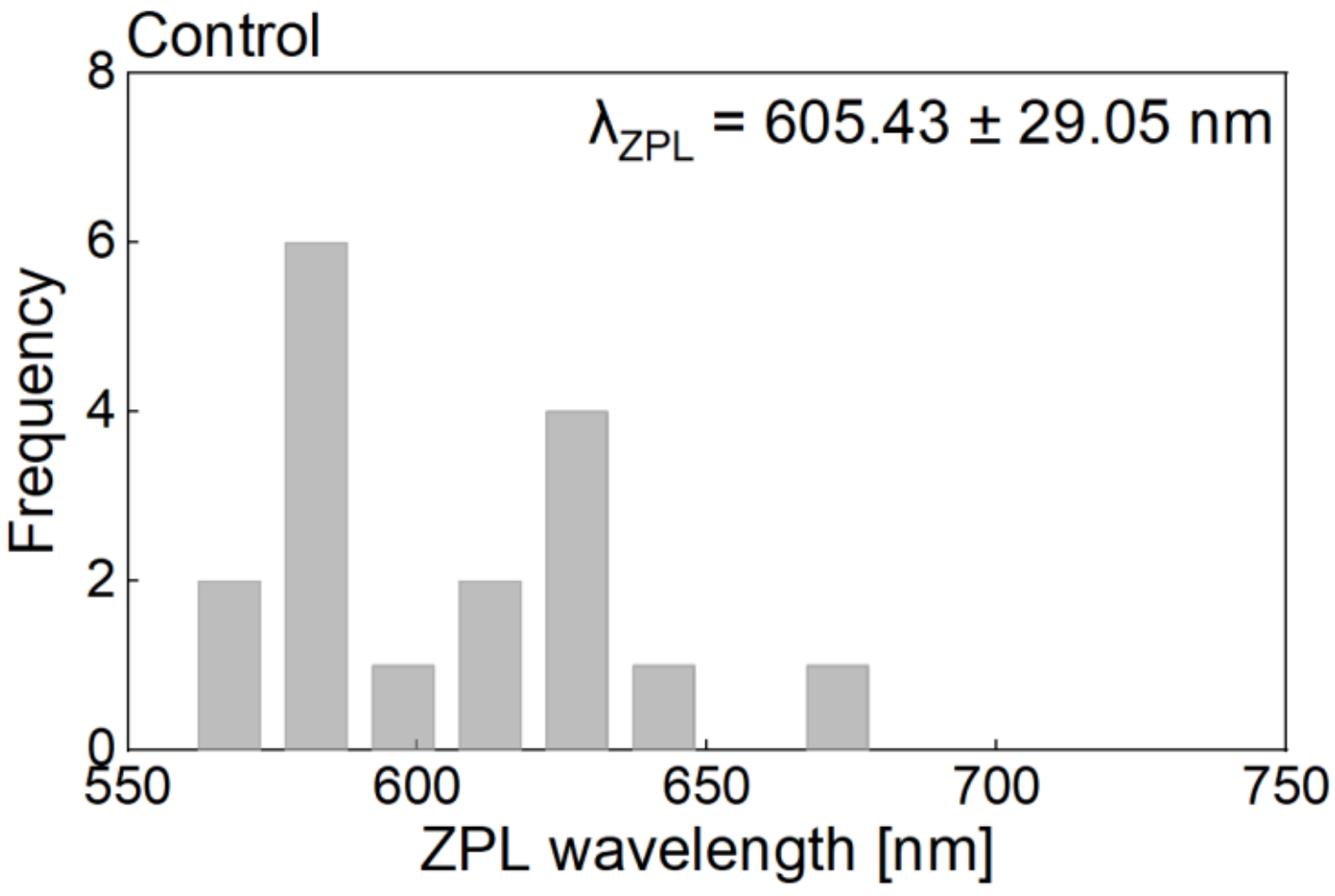

**Figure S8** | ZPL wavelength statistic of emitters identified in non-overlap regions

**- Figure S8** presents the histogram of ZPL wavelength for emitters found in non-overlap areas of samples O1, O2, and O3. The mean ZPL value, 605.43 ± 29.05 nm, is close to that of overlap areas (619.94 ± 46.04 nm). Additionally, the two areas show similar distributions, with most emitters located in the 560-590 nm and around 630 nm ranges, suggesting that the emitters in both regions may share a common origin, potentially associated with C-related, O-related and intrinsic defects.